\documentclass[twocolumn,showpacs,aps,prl,superscriptaddress]{revtex4}

\usepackage{graphicx}
\usepackage{dcolumn}
\usepackage{amsmath}
\usepackage{epsfig}

\RequirePackage{xspace}

\usepackage{relsize}
\def\babar{\mbox{\slshape B\kern-0.1em{\smaller A}\kern-0.1em
    B\kern-0.1em{\smaller A\kern-0.2em R}}}

\def\epem       {\ensuremath{e^+e^-}\xspace}

\def\pip   {\ensuremath{\pi^+}\xspace}
\def\pim   {\ensuremath{\pi^-}\xspace}

\def\pipm  {\ensuremath{\pi^\pm}\xspace}
\def\pimp  {\ensuremath{\pi^\mp}\xspace}

\def\Kbar  {\kern 0.2em\overline{\kern -0.2em K}{}\xspace}

\def\Kz    {\ensuremath{K^0}\xspace}
\def\Kzb   {\ensuremath{\Kbar^0}\xspace}
\def\KzKzb {\ensuremath{\Kz \kern -0.16em \Kzb}\xspace}
\def\Kp    {\ensuremath{K^+}\xspace}
\def\Km    {\ensuremath{K^-}\xspace}
\def\Kpm   {\ensuremath{K^\pm}\xspace}
\def\Kmp   {\ensuremath{K^\mp}\xspace}
\def\KpKm  {\ensuremath{\Kp \kern -0.16em \Km}\xspace}
 
\def\KL    {\ensuremath{K^0_{\scriptscriptstyle L}}\xspace}

\def\Dbar    {\kern 0.2em\overline{\kern -0.2em D}{}\xspace}

\def\Dz      {\ensuremath{D^0}\xspace}
\def\Dzb     {\ensuremath{\Dbar^0}\xspace}
\def\DzDzb   {\ensuremath{\Dz {\kern -0.16em \Dzb}}\xspace}
\def\Dp      {\ensuremath{D^+}\xspace}
\def\Dm      {\ensuremath{D^-}\xspace}

\def\DpDm    {\ensuremath{\Dp {\kern -0.16em \Dm}}\xspace}

\def\Bbar    {\kern 0.18em\overline{\kern -0.18em B}{}\xspace}

\def\BB      {\ensuremath{B\Bbar}\xspace} 
\def\Bz      {\ensuremath{B^0}\xspace}
\def\Bzb     {\ensuremath{\Bbar^0}\xspace}
\def\BzBzb   {\ensuremath{\Bz {\kern -0.16em \Bzb}}\xspace}
\def\Bu      {\ensuremath{B^+}\xspace}
\def\Bub     {\ensuremath{B^-}\xspace}

\def\BpBm    {\ensuremath{\Bu {\kern -0.16em \Bub}}\xspace}

\def\BorBbar    {\kern 0.18em\optbar{\kern -0.18em B}{}\xspace}
\def\DorDbar    {\kern 0.18em\optbar{\kern -0.18em D}{}\xspace}
\def\KorKbar    {\kern 0.18em\optbar{\kern -0.18em K}{}\xspace}

\mathchardef\Upsilon="7107
\def\Y#1S{\ensuremath{\Upsilon{(#1S)}}\xspace}

\mathchardef\Deltares="7101
\mathchardef\Xi="7104
\mathchardef\Lambda="7103
\mathchardef\Sigma="7106
\mathchardef\Omega="710A

\def\Deltabar{\kern 0.25em\overline{\kern -0.25em \Deltares}{}\xspace}
\def\Lbar{\kern 0.2em\overline{\kern -0.2em\Lambda\kern 0.05em}\kern-0.05em{}\xspace}
\def\Sigbar{\kern 0.2em\overline{\kern -0.2em \Sigma}{}\xspace}
\def\Xibar{\kern 0.2em\overline{\kern -0.2em \Xi}{}\xspace}
\def\Obar{\kern 0.2em\overline{\kern -0.2em \Omega}{}\xspace}
\def\Nbar{\kern 0.2em\overline{\kern -0.2em N}{}\xspace}
\def\Xb{\kern 0.2em\overline{\kern -0.2em X}{}\xspace}

\def\mes        {\mbox{$m_{\rm ES}$}\xspace}

\newcommand{\tev}{\ensuremath{\mathrm{\,Te\kern -0.1em V}}\xspace}
\newcommand{\gev}{\ensuremath{\mathrm{\,Ge\kern -0.1em V}}\xspace}
\newcommand{\mev}{\ensuremath{\mathrm{\,Me\kern -0.1em V}}\xspace}
\newcommand{\kev}{\ensuremath{\mathrm{\,ke\kern -0.1em V}}\xspace}
\newcommand{\ev}{\ensuremath{\mathrm{\,e\kern -0.1em V}}\xspace}
\newcommand{\gevc}{\ensuremath{{\mathrm{\,Ge\kern -0.1em V\!/}c}}\xspace}
\newcommand{\mevc}{\ensuremath{{\mathrm{\,Me\kern -0.1em V\!/}c}}\xspace}
\newcommand{\gevcc}{\ensuremath{{\mathrm{\,Ge\kern -0.1em V\!/}c^2}}\xspace}
\newcommand{\mevcc}{\ensuremath{{\mathrm{\,Me\kern -0.1em V\!/}c^2}}\xspace}

\def\mus  {\ensuremath{\rm \,\mus}\xspace}

\def\ps   {\ensuremath{\rm \,ps}\xspace}

\def\mus        {\ensuremath{\,\mu{\rm s}}\xspace}    
\def\ps         {\ensuremath{{\rm \,ps}}\xspace}  

\def\to                 {\ensuremath{\rightarrow}\xspace}

\def\pep2{PEP-II}

\def\gsim{{~\raise.15em\hbox{$>$}\kern-.85em
          \lower.35em\hbox{$\sim$}~}\xspace}
\def\lsim{{~\raise.15em\hbox{$<$}\kern-.85em
          \lower.35em\hbox{$\sim$}~}\xspace}

\def\CP                {\ensuremath{C\!P}\xspace}

\xspace

\newcommand{\jprlBase}       {Phys.\ Rev.\ Lett.\xspace}
\newcommand{\jprBase}        {Phys.\ Rev.\xspace}
\newcommand{\jplBase}        {Phys.\ Lett.\xspace}
\newcommand{\nimBaseA}       {Nucl.\ Instr.\ Meth.\xspace}

\newcommand{\nima}      [1]  {\nimBaseA~A~{\bf #1}}

\newcommand{\plb}       [1]  {\jplBase\ B~{\bf #1}}

\newcommand{\jprl}      [1]  {\jprlBase\ {\bf #1}}
\newcommand{\jprd}      [1]  {\jprBase\ D~{\bf #1}}

\newcommand{\progtp}    [1]  {{Prog.\ Theor.\ Phys.\ {\bf #1}}}

\def\jetset74   {\mbox{\tt Jetset \hspace{-0.5em}7.\hspace{-0.2em}4}\xspace}

\def\fish    {\ensuremath{\cal F}}
\def\akpi  {\ensuremath{{\cal A}_{K\pi}}}

\def\de {\ensuremath{\Delta E}}

\def\figurebox#1#2#3{%
    \def\arg{#3}%
    \ifx\arg\empty
    {\hfill\vbox{\hsize#2\hrule\hbox to #2{\vrule\hfill\vbox to #1{\hsize#2\vfill}\vrule}\hrule}\hfill}%
    \else
    {\hfill\epsfbox{#3}\hfill}%
    \fi}

\long\def\inst#1{\par\nobreak\kern 4pt\nobreak
    {\it #1}\par\vskip 10pt plus 3pt minus 3pt}

\begin{document}

\title{
{
\Large \bf \boldmath 
Direct \CP\ Asymmetry in $\Bz\to\Kp\pim$ Decays}
}

%
\author{B.~Aubert}
\author{R.~Barate}
\author{D.~Boutigny}
\author{F.~Couderc}
\author{J.-M.~Gaillard}
\author{A.~Hicheur}
\author{Y.~Karyotakis}
\author{J.~P.~Lees}
\author{V.~Tisserand}
\author{A.~Zghiche}
\affiliation{Laboratoire de Physique des Particules, F-74941 Annecy-le-Vieux, France }
% ...
\collaboration{The \babar\ Collaboration}
\noaffiliation

\date{\today}
\newpage

\begin{abstract}
We present a measurement of the direct \CP\ asymmetry in the decay 
$\Bz\to\Kp\pim$ using a data sample of $227$ million $\Y4S\to\BB$ decays 
collected with the \babar\ detector at the \pep2\ asymmetric-energy $\epem$ 
collider at SLAC.  We observe a total signal yield of 
$n_{\Km\pip} +  n_{\Kp\pim} = 1606\pm 51$ decays and measure the 
asymmetry 
$\left(n_{\Km\pip}-n_{\Kp\pim}\right)/\left(n_{\Km\pip}+n_{\Kp\pim}\right)
=-0.133\pm 0.030\,(\rm stat) \pm 0.009\,(\rm syst)$.  The probability of
observing such an asymmetry in the absence of direct \CP\ violation is
$1.3\times 10^{-5}$, corresponding to $4.2$ standard deviations.
\end{abstract}


\maketitle

\CP\ violation has been established in processes involving $\Bz$--$\Bzb$ 
oscillations through measurements of the time dependence of neutral-$B$-meson 
decays to final states that include 
charmonium~\cite{BaBarSin2beta,BelleSin2beta}.  Direct \CP\ violation, 
a phenomenon that does not involve particle-antiparticle oscillations,
has been observed in $\KL$ decays~\cite{KlongCP}, where the effect is a 
few parts per million.  In contrast, a large effect is expected in the 
$B$-meson system if \CP\ violation arises from the Kobayashi-Maskawa 
quark-mixing mechanism~\cite{CKM,directCP}.

The Belle Collaboration has reported evidence of direct \CP\ violation
in the decay $\Bz\to\pip\pim$ at the level of 
$3.2\sigma$~\cite{BelleSin2alpha2004}, though this is not confirmed by our 
measurement based on a significantly larger data 
set~\cite{BaBarSin2alpha2004}.
In this Letter we report a measurement of direct \CP\ violation in the decay 
$\Bz\to\Kp\pim$~\cite{cc} at the level of $4.2\sigma$ using a sample of $227$ 
million $\BB$ pairs collected with the \babar\ detector at the SLAC PEP-II $\epem$ 
asymmetric-energy storage ring.  

Direct \CP\ violation is observable as an asymmetry in yields between a
decay and its \CP\ conjugate when at least two contributing amplitudes 
carry different weak and strong phases.  
In the standard model, the decay $\Bz\to\Kp\pim$ occurs through two 
different mechanisms (``penguin'' and ``tree''), which carry different 
weak phases and, in general, different strong phases.  The 
direct \CP-violating asymmetry~\cite{CPTcitation} is defined as
\begin{equation}
\akpi \equiv \frac{n_{\Km\pip}-n_{\Kp\pim}}{n_{\Km\pip}+n_{\Kp\pim}},
\end{equation}
where $n_{\Km\pip}$ and $n_{\Kp\pim}$ are the measured yields for
the two final states.  The charge of the kaon identifies the
flavor of the decaying $B$ meson ($\Bz\to\Kp\pim$, $\Bzb\to\Km\pip$, 
neglecting second-order weak transitions).
The Belle collaboration recently reported a measurement of 
$\akpi = -0.088\pm 0.035\pm 0.013$ using a data sample of $152$ million
$\BB$ pairs~\cite{BelleDirectCP}, which agrees with our
previous result~\cite{BaBarsin2alpha2002}, and with a less-precise measurement from
the CLEO collaboration~\cite{CLEOdirectCP}.

\begin{figure*}[!tb]
\begin{center}
\includegraphics[width=0.32\linewidth]{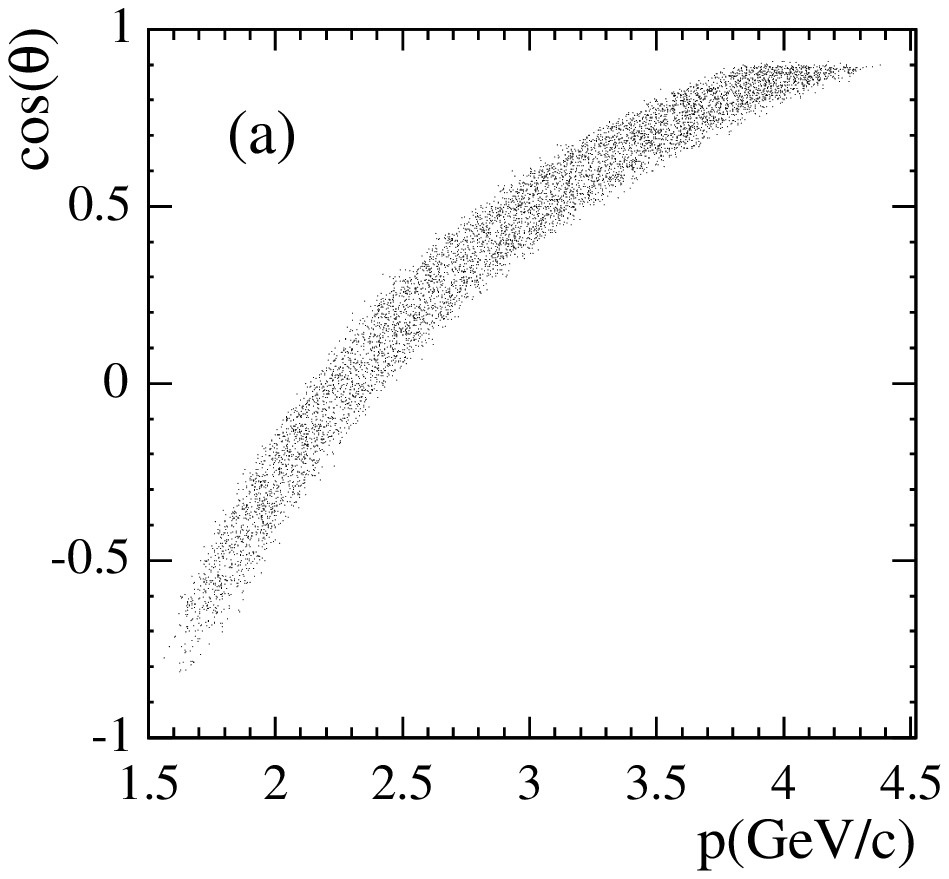}
\includegraphics[width=0.32\linewidth]{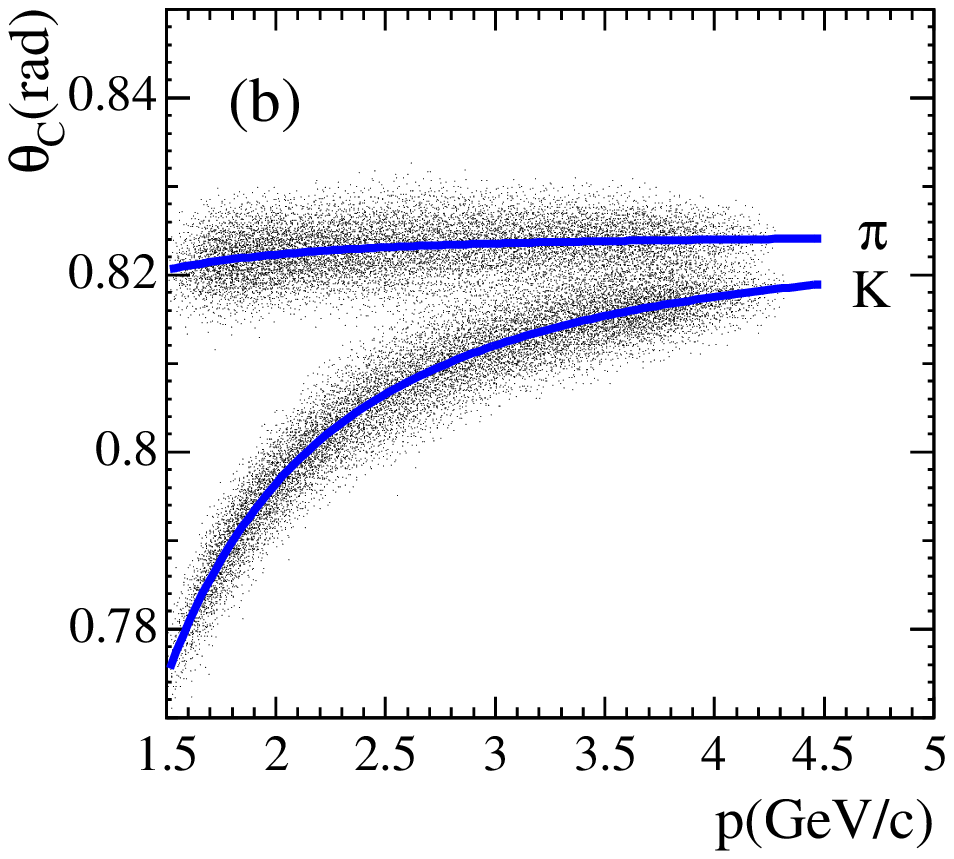}
\includegraphics[width=0.32\linewidth]{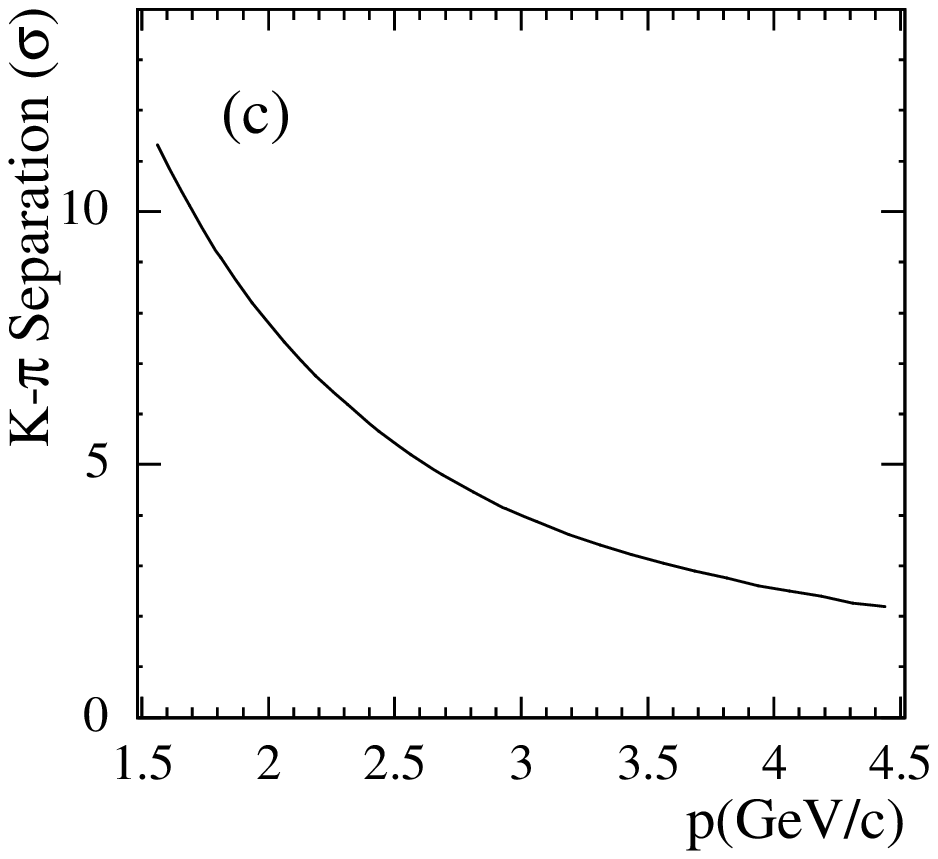}
\end{center}
\caption{(a) Cosine of the polar angle ($\theta$) as a function of laboratory momentum 
for kaons and pions in simulated $\Bz\to\Kp\pim$ decays at \babar.  At a symmetric
$\epem$ collider operating at the \Y4S\ resonance, particles from two-body $B$ decays 
are nearly monoenergetic with $p\sim 2.6\gevc$.  The boost at PEP-II results in an
approximately uniform distribution of laboratory momenta between $1.5$ and $4.5\gevc$, and 
induces a correlation between momentum and polar angle.  (b) The measured Cherenkov
angle for pions (upper band) and kaons (lower band) from $D^{*+}\to\Dz\pip$, $\Dz\to \Km\pip$
decays reconstructed in data.  The curves show the expected angle $\theta_c$
as a function of laboratory momentum, for the $K$ and $\pi$ mass hypothesis.  
(c) The average difference between the expected value of $\theta_c$ for kaons and pions, 
divided by the uncertainty, as a function of momentum.}
\label{fig1}
\end{figure*}

The \babar\ detector is described in detail elsewhere~\cite{BaBar}.
The primary components used in this analysis are a 
charged-particle tracking system consisting of a five-layer silicon 
vertex tracker (SVT) and a 40-layer drift chamber (DCH) surrounded 
by a $1.5$-T solenoidal magnet, an electromagnetic calorimeter 
(EMC) comprising $6580$ CsI(Tl) crystals, and a detector of 
internally reflected Cherenkov light (DIRC), providing $K$--$\pi$ 
separation over the range of laboratory momentum relevant
for this analysis (Fig.~\ref{fig1}).

We reconstruct two-body neutral-$B$ decays from pairs of 
oppositely-charged tracks located within the geometric acceptance of the DIRC
and originating from a common decay point near the interaction region.
We require that each track have an associated Cherenkov-angle ($\theta_c$) 
measured with at least five signal photons detected in the DIRC.  The 
value of $\theta_c$ must agree within $4\sigma$ with either the pion or kaon particle 
hypothesis.  The last requirement efficiently removes events 
containing high-momentum protons.  Electrons are explicitly removed based 
on energy-loss measurements in the SVT and DCH, and on a comparison of the 
track momentum and associated energy deposited in the EMC.  We use the $\theta_c$ 
measurement to separate kaons and pions in a maximum-likelihood fit that
determines signal and background yields corresponding to the four 
distinguishable final states ($\pip\pim$, $\Kp\pim$, $\Km\pip$, $\Kp\Km$).

Signal decays are identified using two kinematic variables: (1) the 
difference $\de$ between the energy of the $B$ candidate
in the $\epem$ center-of-mass (CM) frame and $\sqrt{s}/2$ and (2) the beam-energy 
substituted mass 
$\mes = \sqrt{(s/2 + {\mathbf {p}}_i\cdot {\mathbf {p}}_B)^2/E_i^2- {\mathbf {p}}_B^2}$.
Here, $\sqrt{s}$ is the total CM energy, and the $B$ momentum ${\mathbf {p_B}}$ 
and the four-momentum of the $\epem$ initial state $(E_i, {\mathbf {p_i}})$ are 
defined in the laboratory frame. 
For signal decays, $\de$ and $\mes$ are distributed according to Gaussian
distributions with resolutions of $27\mev$ and $2.6\mevcc$, respectively.  
The distribution of $\mes$ peaks near the $B$ mass for all four particle
combinations.  To simplify the likelihood definition, we reconstruct the 
kinematics of the $B$ candidate using the pion mass for both tracks.
With this choice, $\Bz\to\pip\pim$ decays peak 
near $\de = 0$.  For $B$ decays with one or two kaons in the final state,
the $\de$ peak position is shifted and parameterized as a function of the 
kaon momentum in the laboratory frame.  The average shifts with respect to 
zero are $-45\mev$ and $-91\mev$, respectively.  We require 
$5.20 < \mes < 5.29\gevcc$ and $\left|\de\right|<150\mev$.  The large
sideband region in $\mes$ is used to determine background-shape parameters, 
while the wide range in $\de$ allows us to separate $B$ decays 
to all four final states in the same fit.

We have studied potential backgrounds from higher-multiplicity $B$ decays 
and find them to be negligible in the selected $\de$ region.  
The dominant source of background is the process 
$\epem\to q\bar{q}\; (q=u,d,s,c)$, which produces a distinctive jet-like topology.  
In the CM frame we define the angle $\theta_S$ between the sphericity axis~\cite{sph} 
of the $B$ candidate and the sphericity axis of the remaining particles in the event.  
For background events, $\left|\cos{\theta_S}\right|$ peaks sharply near unity, 
while it is nearly flat for signal decays.  We require $\left|\cos{\theta_S}\right|<0.8$, 
which removes approximately $80\%$ of this background.

The selected sample contains $68030$ events and is composed of
two-body $B$ decays (signal) and combinations of real
kaons and pions produced in $q\bar{q}$ events (background).  
We use an unbinned, extended maximum-likelihood fit to extract yields,
the signal asymmetry ($\akpi$), and the background asymmetry (${\cal A}_{K\pi}^{\rm b}$).  
The fit uses $\mes$, $\de$, $\theta_c$, and the Fisher discriminant ${\cal F}$
described in Ref.~\cite{BaBarsin2alpha2002} to distinguish
signal and background components for each of the four 
$\pip\pim$, $\Kp\pim$, $\Km\pip$, and $\Kp\Km$ combinations.  The 
likelihood for event $j$ is obtained by summing the product of the event 
yield $n_i$ and probability ${\cal P}_i$ over the signal and background 
hypotheses $i$.  The total likelihood for the sample is
\begin{equation}
{\cal L} = \exp{\left(-\sum_{i}n_i\right)}
\prod_{j}\left[\sum_{i}n_i{\cal P}_{i}(\vec{x}_j;\vec{\alpha}_i)\right].
\end{equation}
The probabilities ${\cal P}_i$ are evaluated as the product of 
the probability density functions (PDFs) with parameters $\vec{\alpha_i}$,
for each of the independent variables 
$\vec{x}_j = \left\{\mes, \de, {\cal F}, \theta_c^+,\theta_c^-\right\}$,
where $\theta_c^+$ and $\theta_c^-$ are the Cherenkov angles for the 
positively- and negatively-charged tracks, respectively.  We have verified
that there are no significant correlations between the $\vec{x}_j$.
For both signal and background, the ${\Kpm\pimp}$ yields are parameterized as 
$n_{\Kpm\pimp} = n_{K\pi}\left(1\mp\akpi\right)/2$, and we fit directly
for the total yield $n_{K\pi}$ and the asymmetry $\akpi$.

The $\theta_c$ PDFs are obtained from a sample of approximately $430000$
$D^{*+}\to D^0\pi^+\,(\Dz\to\Km\pip)$ decays reconstructed in data, where 
$\Kmp/\pipm$ tracks are identified through the charge correlation with the
$\pipm$ from the $D^{*\pm}$ decay.  Figure~\ref{fig1}(b) shows the 
measured values of the Cherenkov angle as a function of laboratory momentum
for tracks from the $D^*$ sample, and the expected values for kaons 
$(\theta_c^K)$ and pions $(\theta_c^{\pi})$.  
Figure~\ref{fig1}(c) shows the average $K$--$\pi$ separation, defined as
$\left|\theta_c^K-\theta_c^{\pi}\right|/\sigma_{\theta_c}$, where 
$\sigma_{\theta_c}$ is the average uncertainty for kaon 
and pion tracks for a given momentum. The PDFs are constructed separately for 
$\Kp$, $\Km$, $\pip$, and $\pim$ tracks as a function of momentum and polar angle
using the measured and expected values of $\theta_c$, and its 
uncertainty.  We use the same PDFs for signal and background events.

A total of $21$ parameters are varied in the fit.  Signal and background 
yields and $K\pi$ asymmetries are determined simultaneously with the 
parameters of the signal PDFs for $\mes$ and $\de$, as well as the background 
PDF parameters for $\mes$, $\de$, and $\fish$.  The parameters describing the signal 
$\fish$ distribution are fixed to the values obtained from a large sample of
simulated events, and the parameters of the $\theta_c$ PDFs are fixed to the
values obtained from the $D^*$ study.  The analysis was performed with the 
value of $\akpi$ hidden until the event selection and PDF definitions were 
finalized.

The fitted signal yields are $n_{K\pi} = 1606\pm 51$, 
$n_{\pi\pi} = 467\pm 33$, and $n_{KK} = 3\pm 12$, which are all
consistent with our previously published measurements of the flavor-averaged
branching fractions in these decay modes~\cite{BaBarsin2alpha2002}.  
The direct \CP-violating asymmetry is 
\begin{equation}
\akpi = -0.133 \pm 0.030\,(\rm stat) \pm 0.009\,(\rm syst),
\end{equation}
and the background asymmetry is ${\cal A}_{K\pi}^{\rm b} = 0.001\pm 0.008$.
This result is consistent with, and supersedes, our previous 
measurement~\cite{BaBarsin2alpha2002}.
The correlations of $\akpi$ with ${\cal A}_{K\pi}^{\rm b}$ and $n_{K\pi}$
are $-8\%$ and $+2\%$, respectively.  Correlations with the remaining
free parameters are all $1\%$ or less.

The dominant source of systematic error is the potential difference
between kaons and pions in the dependence of track reconstruction and 
particle identification on the charge of the particle.  To estimate this 
systematic uncertainty, we use the 
statistical uncertainty $(0.008)$ on the measurement of 
${\cal A}_{K\pi}^{\rm b}$ as a conservative systematic error on $\akpi$.  
This background is due to combinations of real kaons 
and pions in the same momentum and polar-angle range as the signal tracks, and 
should have similar sensitivity to a potential bias.  We have also
investigated potential differences in efficiencies for track reconstruction,
and for the requirement of a minimum number of signal photons detected
in the DIRC.  Using the large sample of kaons and pions from the
$D^*$ study, we confirm that the efficiency asymmetries between $\Kp/\Km$
and $\pip/\pim$ are consistent with zero within the small error of the
measurements $(0.002$).  Doubly-Cabibbo-suppressed $\Dz$ decays 
($\Dz\to\Kp\pim$) would produce a bias in the $\theta_C$ PDFs
derived from the $D^*$ sample, but are a negligible effect given the 
current size of the data set.

\begin{figure}[!tb]
\begin{center}
\includegraphics[width=0.9\linewidth]{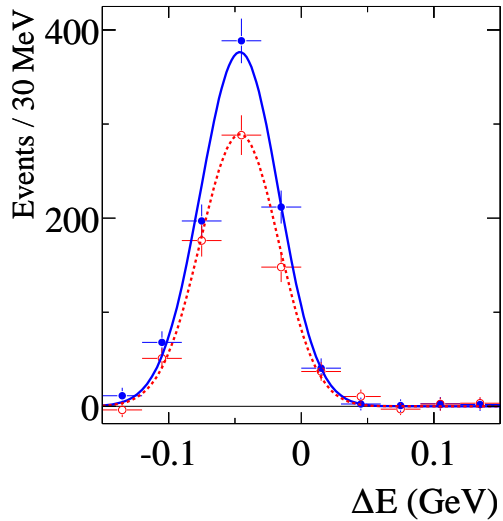}
\caption{Distributions of $\de$ in data (points with error bars) and the 
PDFs (curves) used in the maximum 
likelihood fit for $\Kp\pim$ (solid circles and solid curve) and $\Km\pip$ 
(open circles and dashed curve).  The data are weighted using the background-subtraction 
technique of Ref.~\cite{sPlots} (see text).}
\label{fig2}
\end{center}
\end{figure}

We confirm that we are sensitive to a nonzero value of
$\akpi$ by performing fits on samples of Monte-Carlo-simulated signal 
events, and background events generated directly from the PDF shapes.  
With a generated asymmetry of $-10\%$, the average fitted value in the 
ensemble of events is $\akpi = -0.102\pm 0.002$.  Although the result is 
consistent with the generated value, we take the sum in quadrature of the 
error and the difference with respect to the generated value as a 
systematic uncertainty $(0.003)$.  The systematic errors from uncertainties 
in the distribution of $\fish$ for signal events ($0.001$) and from the parameters 
describing the $\theta_c$ PDFs ($0.001$) are negligible.  The total 
systematic error $(0.009)$ is calculated as the sum in quadrature of the 
individual uncertainties.

Figure~\ref{fig2} shows background-subtracted distributions of $\de$
for signal $\Kp\pim$ and $\Km\pip$ decays.  The subtraction is performed using 
the technique described in Ref.~\cite{sPlots}, where each event is given a
statistical weight that depends on the PDFs and covariance matrix from a
fit excluding the variable being plotted.  The resulting distribution is
normalized to the signal yield and its shape can be compared with
the PDF we use in the full fit.  We see no evidence of an enhancement
near $\de =0$, which could arise from significant contamination of 
$\Bz\to\pip\pim$ decays due to imperfect parameterizations of the 
$\theta_c$ PDFs.

\begin{figure}[!tb]
\begin{center}
\includegraphics[width=0.9\linewidth]{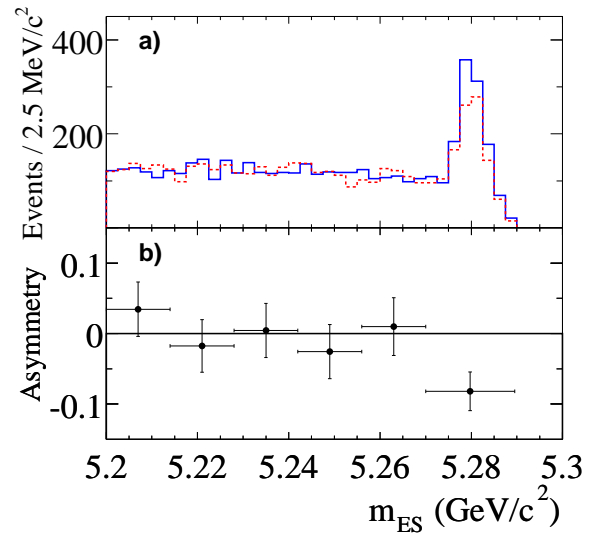}
\caption{(a) Distribution of $\mes$ enhanced in $\Kp\pim$ (solid histogram) 
and $\Km\pip$ (dashed histogram). (b) Asymmetry $\akpi$ calculated for
ranges of $\mes$.  The asymmetry in the highest $\mes$ bin is somewhat 
diluted by the presence of background.}
\label{fig3}
\end{center}
\end{figure}

As a further consistency check on the fit result, in Fig.~\ref{fig3}(a) we show 
distributions of $\mes$ for samples enhanced in signal $K\pi$ decays using 
probability ratios based on the PDFs for $\de$, $\fish$, and $\theta_c$.
The efficiency of the selection is approximately $80\%$ for signal $K\pi$
decays, while the contamination from $\Bz\to\pip\pim$ is less than $2\%$.
Fig.~\ref{fig3}(b) shows the resulting distribution of $\akpi$ as a function
of $\mes$.

A number of consistency checks are performed to validate the result.  We generate
and fit a large set of pseudo-experiments, where the variables $\vec{x}_j$
for each event are generated randomly from the PDFs, and confirm that
the value of $\akpi$ is intrinsically unbiased.  To check for a potential effect
from $K$--$\pi$ misidentification, we fit the subsample of events (less than half) 
where both tracks have laboratory momentum less than $3.5\gevc$.  The $K$--$\pi$
separation for all tracks in this sample is greater than $3\sigma$, and we 
find $\akpi = -0.151\pm 0.047$.  
We perform a $\Bz$--$\Bzb$ mixing analysis on the full two-body sample 
using $B$-flavor identification and decay-time information as described in 
Ref.~\cite{BaBarsin2alpha2002}.  From this fit, we simultaneously determine 
the $B$ lifetime $\tau_B = 1.60\pm 0.04\ps$, $\Bz$--$\Bzb$ mixing frequency 
$\Delta m_d=0.523\pm 0.028\,{\rm ps}^{-1}$, and $\akpi = -0.126\pm 0.029$, 
where the errors are statistical only.  The values of $\tau_B$ and 
$\Delta m_d$ are consistent with the world averages~\cite{PDG2004}, and 
$\akpi$ is consistent with the nominal fit, demonstrating that the signal
events have the expected time evolution.  Finally, we divide the full 
sample into the approximate period in which the data were recorded 
(Table~\ref{table}).  We find $\akpi<0$ and a background asymmetry consistent 
with zero in each data set.

\begin{table}[!tb]
\begin{center}
\caption{The signal yield $n_{K\pi}$, and signal ($\akpi$) and
background (${\cal A}_{K\pi}^{\rm b}$) asymmetries measured in different
data-taking periods.  The number of $\BB$ pairs $N_{\BB}$ (in millions) for each 
data set is also given.}
\label{table}
\begin{ruledtabular}
\begin{tabular}{cccccc} 
Sample        & $N_{\BB}$ & $n_{K\pi}$ & $\akpi$   & ${\cal A}_{K\pi}^{\rm b}$\\ 
\hline
$1999$--$2001$ & $21.1$    & $142\pm 15$ & $-0.240\pm 0.102$ & $0.006\pm 0.026$ \\
$2002$         & $66.4$    & $479\pm 27$ & $-0.102\pm 0.055$ & $-0.008\pm 0.015$\\
$2003$         & $34.1$    & $241\pm 19$ & $-0.109\pm 0.079$ & $0.007\pm 0.021$ \\
$2004$         & $104.9$   & $743\pm 33$ & $-0.142\pm 0.044$ & $0.004\pm 0.012$ \\
\end{tabular}
\end{ruledtabular}
\end{center}
\end{table}

The statistical significance of the measurement ($4.3\sigma$) is 
computed by taking the square root of the change in $2\ln{\cal L}$ when 
$\akpi$ is fixed to zero.  If we include the systematic error by 
summing in quadrature with the statistical uncertainty, the significance 
is $4.2\sigma$, and the probability of obtaining a negative asymmetry
of this magnitude or larger in the absence of \CP\ violation is 
$1.3\times 10^{-5}$.  We conclude that the measurement of 
$\akpi = -0.133\pm 0.030\,(\rm stat) \pm 0.009\,(\rm syst)$ reported here 
represents compelling evidence for direct \CP\ violation in the $B^0$-meson 
system.

\par

We are grateful for the excellent luminosity and machine conditions
provided by our \pep2\ colleagues, 
and for the substantial dedicated effort from
the computing organizations that support \babar.
The collaborating institutions wish to thank 
SLAC for its support and kind hospitality. 
This work is supported by
DOE
and NSF (USA),
NSERC (Canada),
IHEP (China),
CEA and
CNRS-IN2P3
(France),
BMBF and DFG
(Germany),
INFN (Italy),
FOM (The Netherlands),
NFR (Norway),
MIST (Russia), and
PPARC (United Kingdom). 
Individuals have received support from CONACyT (Mexico), A.~P.~Sloan Foundation, 
Research Corporation,
and Alexander von Humboldt Foundation.


\begin{thebibliography}{99}
\bibitem{cc}
Unless otherwise stated, charge conjugate modes are included implicitly.
\end{thebibliography}
\end{document}